\documentclass[axodraw,fleqn,twoside]{article}
\usepackage{espcrc2}
\usepackage{axodraw}

\usepackage{epsfig}
\usepackage{graphicx}

\usepackage[figuresright]{rotating}

\newcommand{\AmS}{{\protect\the\textfont2 A\kern-.1667em\lower.5ex\hbox{M}\kern-.125emS}}

\title{Light-Front model of transition form-factors in heavy
meson decay } 

\author{J.~P.~B.~C.~de~Melo \address[LFTC]{
Laborat\'orio de F\'\i sica Te\'orica e Computa\c c\~ao Cient\'\i
fica, LFTC, Universidade Cruzeiro do Sul \\ 01506-000, S\~ao
Paulo, SP, Brazil} \address[IFT] {Instituto de F\'\i sica
Te\'orica, Universidade Estadual Paulista, IFT-UNESP, Rua Dr.
Bento Teobaldo Ferraz, 271 Bl. II  Barra Funda, 01140-070, S\~ao
Paulo, SP, Brazil}, T.~Frederico\address[ITA]{Dep. de F\'isica,
Instituto Tecnol\'ogico de Aeron\'autica, 12.228-900, S\~ao Jos\'e
dos Campos, Brazil} }

\begin{document}

\begin{abstract}
Electroweak transition form factors of heavy meson decays are
important ingredients in the extraction of the
Cabibbo-Kobayashi-Maskawa~(CKM) matrix elements from experimental
data. In this work, within a light-front framework, we calculate
electroweak transition form factor for the semileptonic decay of
$D$ mesons into a pion or a kaon. The model results underestimate
in both cases the new data of CLEO for the larger momentum
transfers accessible in the experiment. We discuss possible
reasons for that in order to improve the model. \vspace{1pc}
\end{abstract}

\maketitle

\section{Introduction}

Charmed mesons inaugurated a new era in particle physics. The
discovery of the first heavy meson was an evidence in favor of the
standard model as giving the correct description of low energy
particle physics. The new quark was confirmed experimentally (see
\cite{PDG}) in a meson quark-antiquark state, i.e., in a
heavy-light quark system.  Since then, the study of semileptonic
decays gained importance, mainly in the last years, as these
decays constitute a laboratory to test the standard model
predictions and to get information about the CKM~\cite{CKM} matrix
elements. These matrix elements, $V_{cq}$, can be extracted from
the experimental data of D decay. However, theoretical inputs are
necessary in order to interpret the data from the decay
experiments as obtained by BES~\cite{BES},
Babar~\cite{Widham},~Belle~\cite{Aubert},    and
CLEO~\cite{CLEO2009,Ge} collaborations.


The QCD nonperturbative effects in the semileptonic decay are
modelled with transitions form factors that have been calculated
with lattice QCD~\cite{Aubin}, QCD sum rules~\cite{Simula2007} and
light-front constituent quark
models~\cite{Grach1996,Bruno2009,Choi2009}. This motivates us to
study the semileptonic decay of D mesons  within a light-front
covariant model (LFCM) proposed in ref.~\cite{LC2008} and applied
to calculate the pion, kaon and D electromagnetic form factors.

In general, light-front models of hadrons make possible the
calculation of many observables, with the advantage of working at
the amplitude level obtained directly from matrix elements taken
between hadron light-front wave functions. The  valence component of
the  wave function allows a direct interpretation of the relevant
model parameters, being also an useful concept when analytical forms
of the hadron Bethe-Salpeter amplitude are the starting point to
calculate the physical quantities. Applications of light-front
models have been performed in many cases like in the calculation of
pseudoscalar meson decay constants~\cite{Salcedo2006},
electromagnetic form factors of the $\rho$-meson~\cite{Pacheco97},
pion~\cite{Pacheco99,Pacheco2002} and kaon ~\cite{Fabiano07}. In
particular constituent light-front quark models (CLFQM), are
successfully in describing  the pion electromagnetic form factor
compared with the new experimental data ~\cite{piondata} (see e.g.
\cite{Salme2006} and references therein).

One important theoretical aspect for the calculation of
electroweak form factors performed within the light-front
approach, is the inclusion of the non-valence contribution besides
the valence one when the Drell-Yan condition is not respected, as
in the case of the computation of time-like from elastic and
transition factors (see a discussion in ~\cite{Salme2006}). The
nonvalence contribution to the electroweak matrix elements of the
current are associated with Fock-components of the wave function
beyond the valence, that can be as well attributed to two-body
currents acting on the valence wave function \cite{crji,adnei}.
Notably, in some particular light-front models, the nonvalence
contribution should be included in the calculation of
electromagnetic current matrix elements to satisfy the
requirements of covariance in the Drell-Yan
frame~\cite{Pacheco99}. Moreover, in calculations of electroweak
observables within light-front frameworks, apart from the frame
used, it is important to note that the contribution of nonvalence
terms depends on which component of current is employed to extract
the form factors (see ~\cite{Pacheco99,Bakker2001} for details).

Recently, we have proposed an analytical covariant model of the
Bethe-Salpeter amplitude of mesons~\cite{LC2008} and we have used a
Wick rotation of the light-front energy in the Mandelstam formula to
calculate electromagnetic form factors~(LFCM). The Wick rotation
applied to the minus component of the loop-momentum allows a direct
four-dimensional integration avoiding the singularities of the
Bethe-Salpeter amplitudes in Minkowski space in the calculation of
the observables. The computation takes care implicitly,  all
possible zero-modes or nonvalence contributions to the matrix
elements of the electromagnetic current. We have used the above
model and calculational strategy to obtain electromagnetic form
factors in the space-like regime and also decay
constants of light and heavy pseudoscalar mesons. 

Once elastic form factors of pseudoscalars have been calculated in
the model, it is worthwhile to pursue the investigation of
transition form factors in heavy meson decay. The semileptonic
decays of the heavy mesons offer an useful laboratory to test
theoretical models of the corresponding transition form factors in
the weak sector of the standard model. Here, the covariant
Bethe-Salpeter vertex model and computational strategy using
Wick-rotation on the light-front are applied to calculate the
electroweak transition form factors, $f_+(q^2)$ and $f_{-}(q^2)$
for $D$ meson decays into a pion or a kaon. In sect. 2, we
describe the model and give their parameters. In sect. 3, the form
factors and kinematics for the decay are presented. The numerical
results and conclusions are presented in sect. 4.

\section{The vertex model }

The vertex model of the Bethe-Salpeter amplitude for the
meson-$q-\bar{q}$ proposed in  reference~\cite{LC2008} is given by
the following expression:
\begin{eqnarray}
 \Lambda_M(k,p)
\hspace{-0.6cm} & & =
 (k^2-m_q^2)((p-k)^2-m_Q^2) \\ \nonumber
 & &  \hspace{-1.0cm} \times
\frac{\Gamma_M }{(k^2-\lambda^2_M + \imath \epsilon)^n~((p-k)^2-\lambda^2_M +
\imath \epsilon)^n}.
\label{vertex}
\end{eqnarray}
Here, $\lambda_M$ is the scale associated with the meson light-front
valence wave function and $n$ is the power of the regulator.  The
masses of the constituent quarks within the meson state are $m_q$
and $m_Q$. In the vertex function, the factors in the numerator
avoid the cut due $q\bar{q}$ scattering if $m_q+m_Q~<~M_{h}$. In
order to confine the quarks, the regulator scale $\lambda_M$ is
chosen such that ~$\lambda_M > M_{h}/2$, where $M_{h} $ is the meson
mass. This constraint does not allow the meson decay into particles
of mass $\lambda_M$ .

\begin{table}[h]
\begin{tabular}{|l|l|l|l|l|l|}
\hline
 & Model & $r_{ps}$  & $\lambda_M$ & $m_{q}$&  $m_Q$  \\
 & & fm & MeV & MeV & MeV
\\
\hline
$\pi^\pm$   & P3  & 0.296   & 3564 & 210 & 210  \\
       & P4  & 0.350    & 2995 & 215 & 215 \\
       & P5  & 0.392   & 2666  & 220 & 220  \\
\hline
$K^\pm$ & K1 & 0.344 &  3056 & 203 & 327 \\
     & K2 & 0.349 &  3002  & 204 & 328 \\
     & K3 & 0.380 &  2744 & 210 & 334 \\
     & K4 & 0.400 & 2593 & 215 & 339 \\
     & K5 & 0.417  & 2478 & 220 & 344 \\
\hline
 $D^+$ & D1 & 0.341 & 2163  & 203  & 1442 \\
       & D2 & 0.343 & 2162 & 204  & 1443 \\
       & D3 & 0.344 & 2153 & 210  & 1449 \\
       & D4 & 0.345 & 2146  & 215  & 1454  \\
       & D5 & 0.346 & 2138 & 220 & 1459 \\
\hline
\end{tabular}
\caption{ LFCM parameters for the pion, kaon and $D^+$. For each
meson the range parameter $\lambda_M$ of the Bethe-Salpeter vertex
(\ref{vertex}) with $n=10$ is fitted to the corresponding
pseudoscalar decay constant for the given constituent quark
masses. The values used are $f_\pi=94$~MeV, $f_K=113$~MeV and
$f_D=144.5$~MeV \cite{PDG}. The different constituent quark masses
originate the parameterizations labelled by P, K and D, for the
pion, kaon and D, respectively.} \label{table:1} \vspace {-1cm}
\end{table}

The general matrix elements for the electromagnetic current, are
written as a tree-point function, using the Mandelstam formula,
given in the equation below:
\begin{equation}
\left\langle p^{\prime }\left\vert J_{q}^{\mu }\left( q^{2}\right)
\right\vert p\right\rangle
\hspace{-0.1cm}
 = \hspace{-0.1cm}
\frac{N_{c}}{\left( 2\pi \right) ^{4}}
\hspace{-0.1cm}
\int \hspace{-0.1cm}
d^{4}k~Tr \left[ {\cal O}  \right] F(k,p,p^{\prime}).
\end{equation}
where the operator ${\cal O}$  is
\begin{eqnarray*}
{\cal O}=\left[ S_{F}\left(
 k-p^{\prime }\right)
 \gamma^{\mu }S_F(k-p) S_{F}\left( k\right) \right] .
\end{eqnarray*}
$N_c$$S_F(p)$ is the Feynman propagator for constituent quarks and
$$F(k,p,p^{\prime})=\Lambda_{M^{\prime }}\left( k,p^{\prime }\right)
\Lambda _{M}\left( k,p\right).$$

The electromagnetic form factor for pseudoscalar mesons is obtained
from the matrix element of the conserved vector current as:
\begin{equation}
\left\langle p^{\prime }\left\vert J_{q}^{\mu }\left( q^{2}\right)
\right\vert p\right\rangle = \left( p+p^{\prime}\right)^\mu
F_{PS}^{em}(q^2),
\end{equation}
with the choice of plus component of the current, $J^{+}=J^0+J^3$,
that has the minimal contribution from the instantaneous terms of
the Feynman propagator of the constituent quark. Generally, this
"good" component of the current is protected against the
contribution from zero-modes in the Drell-Yan frame, although a
careful analysis has to be performed for each choice of the vertex,
when the integration over the minus component of the loop-momentum
is done analytically \cite{Pacheco99,Bakker2001}. The normalization
condition of the vertex function is such that $F_{PS}^{em}(0)=1$.

The weak decay constant of the pion is defined by the matrix element
of the axial-current:
\begin{equation}
\left\langle 0 \left\vert A^{\mu}(0) \right\vert p\right\rangle =
\imath \sqrt{2} f_{PS} p^{\mu} ,
\end{equation}
where $ A^{\mu}=\bar{q}(x)\gamma^{\mu}\gamma^{5}
\frac{\tau}{2}q(x)$, with analogous expression for the kaon and D.

\begin{figure}[thb]
\vspace{-.8cm}
\includegraphics[angle=0, scale=.34]{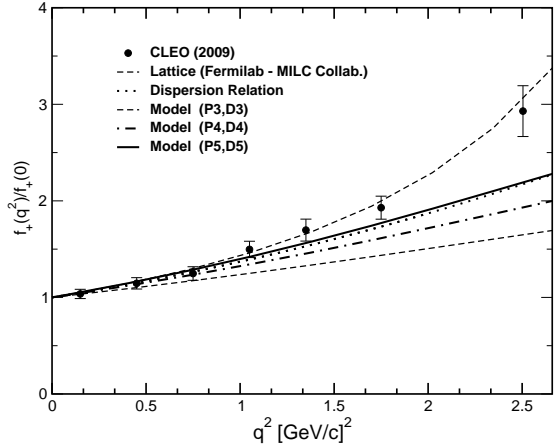}
\vspace{-1.0003cm} \caption{Transition  form factor $f_+(q^2)$
normalized to $f_+(0)=1$ for $D\rightarrow \pi l \nu$.
Calculations with the LFCM model compared to results from lattice
~\cite{MILC} and dispersion relation
calculations~\cite{BRUNO2008}. 
} \label{lcfig1}\vspace{-0.5cm}
\end{figure}
\begin{figure}[hbt]
\vspace{-0.4cm}
\includegraphics[angle=0, scale=.34]{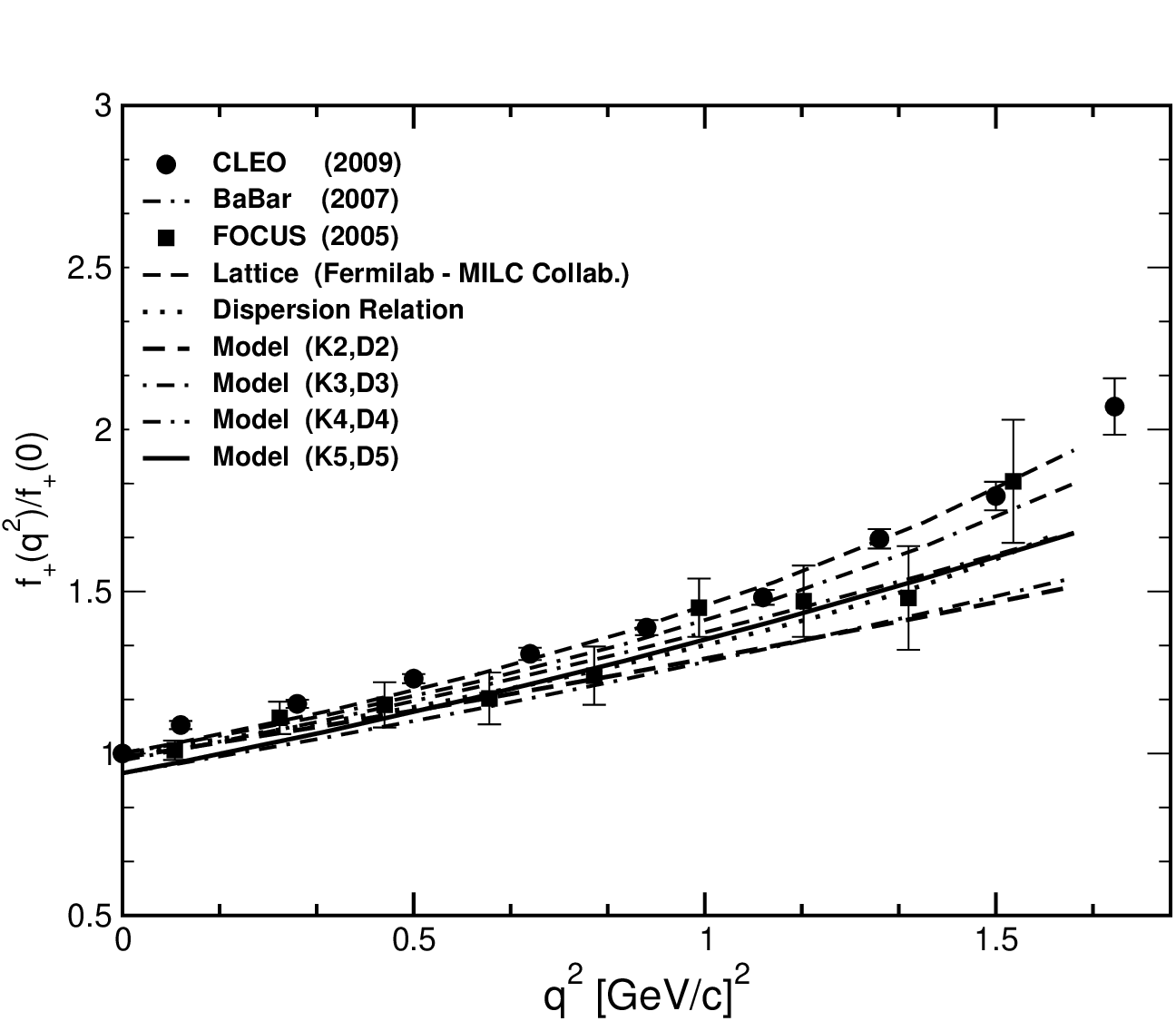}
\vspace{-1.3cm} \caption{Transition  form factor $f_+(q^2)$
normalized to $f_+(0)=1$ for $D\rightarrow K l \nu$. Calculations
with the LFCM model compared to results from lattice ~\cite{MILC}
and dispersion relation calculations~\cite{BRUNO2008}.}
\label{lcfig2} \vspace{-.6cm}
\end{figure}

The final expression for decay constant of the pseudoscalar mesons
is given by
\begin{equation}
 \imath p^\mu f_{PS}=
 N_c \int \frac{dk^4}{(2 \pi)^4} Tr\left[ {\cal G^\mu } \right]
\Lambda_{M}(k,p), \label{decaycte}
\end{equation}
where $PS=(\pi,$K,D) and
$$
{\cal G}=Tr\left[ \gamma^\mu \gamma^5  S_F(k) \gamma^5 S_F(k-p)
\right] \ .
$$
The plus component of the axial current is adopted as discussed
above for the electromagnetic current. The numerical integration
in the Euclidean minus moment component allows one to use, equally
well, any other component.

The vertex model parameters for the pion, kaon and D are obtained
from the decay constants for given constituent quark masses as
presented in table 1. Our choice of constituent quark masses follows
the suggestion that the chiral symmetry breaking mechanism produces
the light constituent quark mass, and the other quarks like $s$ and
$c$ have in addition the current quark mass. In our case we use the
current quark masses of 124 MeV and 1239 MeV for $s$ and $c$,
respectively. We follow the reasonings presented in \cite{suisso},
and the current quark masses are within the accepted range of values
(see \cite{PDG}).

\section{Transition Form Factors}


The transition form factors of the weak vector current investigated
here appears in the semileptonic processes  $D\rightarrow \pi l \nu$
and $D_{s}\rightarrow K l \nu$ decays. In particular, the leptonic
decay of $D_{s}$ allows to extract the decay constant. In the
semileptonic decay of a heavy pseudoscalar meson to a light one, the
nonperturbative dynamics of QCD appears through the matrix elements
of the weak vector current parameterized in terms of two transitions
form factors, $f_+(q^2)$ and $f_{-}(q^2)$:
$$
 \langle \pi (K)|V^{\mu}|D (D_s) \rangle  =
(P^\prime+P)^{\mu} f_+(q^2) + f_-(q^2)q^{\mu} , \label{traffactor}
$$
where $V^{\mu}=\bar{q} \gamma_{\mu} c$, with $q=(d,s)$, and
 $q^\mu=(P^\prime-P)^{\mu}$. $P^\mu$ stands for the four-momentum of the D
 meson and ${P^\prime}^\mu$ for the light pseudoscalar in the final
 state.  The matrix element of the weak vector current can be written
also as:
\begin{eqnarray}
& &  \langle \pi (K)|V^{\mu}|D (D_s) \rangle= \nonumber
\\ \nonumber
& & f_+(q^2)
  \bigg[ (P^\prime+P)^{\mu}
  -\frac{M^2_{D(D_s)}-M^2_{\pi(K)}}{q^2} q^{\mu} \bigg] \\
  & & + f_0(q^2)  \frac{M^2_{D(D_s)}-M^2_{\pi(K)}}{q^2} q^{\mu},
  \end{eqnarray}
when the momentum transfer vanishes $f_{+}(0) = f_0 (0)$.
In the calculation of transition form factors, we use the plus
component of the weak vector current.

The reference frame used in the case of the elastic
 electromagnetic form factors is  defined
 by ~\cite{Salme2006,LC2008}:
$P^{+}  =  {P^\prime}^{+}+q^{+}$, $q^+ \geq 0 $,
$\vec{P}^\prime_\perp =  \vec{P}_\perp=0$ and
${P^\prime}^2=P^2=M^2_{h}$. For the decay process,  we use the
following reference frame: $P^{+}  =  {P^\prime}^{+}-q^{+}$, $q^+
< 0$ and $\vec{P}^\prime_\perp =  \vec{P}_\perp=0$. The numerical
integration of the matrix elements is done with the Wick rotation
of the minus momentum as formulated in \cite{LC2008}.

\section{Results and Conclusion}

In the LFCM models used use $n=10$, we have chosen the range
parameters $\lambda_M$ by  fitting the experimental pseudoscalar
decay constants, the are presented in the table 1, together with
the charge radius. The experimental pion charge radius is 0.672
fm\cite{piondata} much larger than our results, that comes as a
consequence of the regulator power $n=10$. The same is observed
for the kaon which has an experimental charge radius of 0.56 fm
\cite{piondata}.

\begin{table}[htb]
\vspace{-0.6cm}
\begin{tabular}{|l|l|l|}
\hline
 Source & $f^{D \rightarrow K}_+(0) $ & $f^{D \rightarrow \pi}_+(0)$  \\
\hline
 BES~\cite{BES}      & 0.780$\pm$ & 0.730$\pm$\\
 Belle~\cite{Widham} & 0.695$\pm$0.023   & 0.624$\pm$0.036 \\
BaBar~\cite{Aubert}  & 0.728$\pm$  & - \\
 CLEO~\cite{CLEO2009} & 0.739$\pm$0.009 & 0.665$\pm$0.019 \\
CLEO-c~\cite{Ge}  & 0.763$\pm$  & 0.628$\pm$ \\
 LQCD~\cite{Aubin}  &  0.733$\pm$ & 0.643$\pm$ \\
\hline
Model & $f^{D \rightarrow K}_+(0) $ & $f^{D \rightarrow \pi}_+(0)$  \\
\hline
(K1,D1,~--~)& 0.729& -\\
(K2,D2,~--~)& 0.772 & -\\
(K3,D3,P3)&0.913 & 0.491 \\
(K4,D4,P4)&1.020 & 0.739\\
(K5,D5,P5)&1.100 & 0.936\\
\hline
\end{tabular}
\caption{Experimental, lattice and model transition form factor
results for $q^2=0$. } \label{table:2} \vspace{-0.6cm}
\end{table}

We should observe that smaller power in the vertex function
$\Lambda_M(k,p)$ give a good fit for low energy observables
calculated with this model~\cite{LC2008} and describe   well the
electromagnetic form factors of the pion and kaon~\cite{LC2008}.
But, in the case of the transition electromagnetic form factors we
need the higher power $n$, in order to reproduce the experimental
transition form factor $f_{+}(q^2)$ in the low momentum region
(see figures 1 and 2). While the slope near $q^2=0$ seems
reasonable, the calculation underestimate the experimental data
for large momentum transfers for the range of constituent quark
masses used. The effect of a light quark mass variation of 10\% is
seen mainly at large momentum transfers.

In table~\ref{table:2}, we show transition form factors $f_{+}(0)$
for $D$  and $D_s$ compared to the experimental
data~\cite{CLEO2009,Widham,Aubert,Ge,BES} and to lattice
calculation~\cite{MILC}. The sensitivity to the constituent quark
mass is strong, and no simultaneous fit of $D$ and $D_s$
transition form factor at zero momentum transfers was achieved by
the model. The small sizes of the pion and kaon are presumably
responsible for the missing curvature demanded at large momentum
transfers by the data. The large power implied in a high value of
the scale $\lambda_M$ of about 3 GeV for the light mesons, to
compensate for the power $n=10$ while fitting the decay constant.
This hard scale as seen in table 1 shrinks both pion and kaon, as
well as the D meson. The overlap between the wave function
represented by $f^+(0)$ and the slope of the transition form
factor seems not that sensitive to these hard scales but to the
physical decay constants. Therefore, the vertex has to have a much
softer dependence in momentum, and a smaller power should be
required for a better fitting of these observables. Indeed the
expected behavior of the vertex function with momentum should have
a power about $n=2$, as expected from the one-gluon exchange and
the monopole decay of the pion electromagnetic form factor.
Therefore, it is expected that even the $D$ meson should be
somewhat larger than the present calculations indicate of a charge
radius of 0.34 fm, in order to accommodate a reasonable fit to the
transition form factors.

We thank  FAPESP and CNPq of Brazil for the partial financial
support. 

\end{document}